\documentclass{article}

\usepackage{url}
\usepackage{amsthm,graphicx}

\newcommand{\coll}[1]{\url{#1}}

\usepackage{verbatim}

\newcommand{\notaestesa}[2]{%
  {\sffamily {\bfseries #1}{\footnotesize #2}}%
  \marginpar{\framebox{\Large *}}%
}
\begin{document}

\title{PIntron: a Fast Method for Gene Structure Prediction  via Maximal
  Pairings of a Pattern and a Text}
\author{%
  Paola Bonizzoni\thanks{DISCo, Univ.  Milano--Bicocca, viale Sarca 336, 20126 -
  Milano. Italy} \thanks{corresponding author}
  \and
  Gianluca Della Vedova\thanks{Dip. Statistica, Univ. Milano--Bicocca, via
    Bicocca degli Arcimboldi 8, 20126 - Milano. Italy.}
  \and
  Yuri Pirola$^*$
  \and
  Raffaella Rizzi$^{*}$}

\newtheorem{Theorem}{Theorem}[section]
\newtheorem{Lemma}[Theorem]{Lemma}
\newtheorem{Proposition}[Theorem]{Proposition}
\newtheorem{Observation}[Theorem]{Observation}
\newtheorem{Corollary}[Theorem]{Corollary}
\newtheorem{Claim}[Theorem]{Claim}
\newtheorem{Property}[Theorem]{Property}
\theoremstyle{definition}
\newtheorem{Definition}[Theorem]{Definition}
\newtheorem{Problem}{Problem}
\newtheorem{Example}{Example}[section]

\newcommand{\mathfrc}[1]{\text{\fontfamily{frc}\selectfont#1}}

\newcommand{\inputdata}[1]{\noindent \emph{Input: }#1\\*}
\newcommand{\outputdata}[1]{\noindent \emph{Output: }#1\\}
\newcommand{\ie}{i.e.~}
\newcommand{\st}{s.t.~}
\newcommand{\wrt}{w.r.t.~}
\newcommand{\emb}{\ensuremath{\varepsilon}}
\newcommand{\embset}{\ensuremath{\mathcal{E}}}
\newcommand{\Path}{\ensuremath{\mathcal{P}}}
\newcommand{\nil}{\ensuremath{\bot}}
\newcommand{\card}[1]{\ensuremath{|#1|}}
\renewcommand{\l}{\ensuremath{\ell}}
\renewcommand{\emptyset}{\ensuremath{\varnothing}}

\maketitle
\begin{abstract}
Current computational  methods for exon-intron structure
prediction from a cluster of transcript (EST, mRNA) data do not
exhibit the time and space efficiency  necessary to process large
clusters of over than 20,000 ESTs and  genes longer than 1Mb. Guaranteeing
both accuracy and efficiency
seems to be  a computational goal quite far to be achieved, since
accuracy is strictly related to exploiting the inherent redundancy
of information present in a large cluster.

We propose a fast method for the problem that combines two  ideas:
a novel   algorithm of proved small time complexity for computing
spliced alignments of  a transcript   against a genome, and  an
efficient  algorithm that exploits the inherent
redundancy of information in a cluster of transcripts to select,
among all possible factorizations of EST sequences, those allowing
to infer splice site junctions that are highly confirmed by the
input data. The  EST alignment procedure is based on the
construction of \emph{maximal embeddings} that are sequences
obtained  from  paths of a graph structure, called
Embedding Graph, whose vertices are the \emph{maximal pairings} of
a genomic sequence $T$ and an EST $P$. The
procedure runs  in time linear in the size of  $P$, $T$ and  of
the output.

 PIntron, the software tool implementing our methodology, is able
 to process in a few seconds some critical genes  that are not manageable by
 other gene structure
prediction tools. At the same time, PIntron  exhibits high accuracy (sensitivity and
specificity) when compared with ENCODE data.

Detailed experimental data, additional results  and
PIntron  software are
available
at \coll{http://www.algolab.eu/PIntron}.
\end{abstract}

\section{Introduction}
\label{sec:spl:intro}

A key step in the post-transcriptional modification process is
called \emph{splicing} and consists in the excision of the
intronic regions of the premature mRNA (pre-mRNA) while the exonic
regions are then reconnected to re-form a single continuous
molecule, the mature mRNA. A complex regulatory system mediates
the splicing process which, under different conditions, may
produce \emph{alternative} mature mRNAs (also called transcript
isoforms) starting from a single pre-mRNA molecule. Alternative
Splicing (AS), \ie the production of alternative transcripts from
the same gene, is the main mechanism responsible for the expansion
of the transcriptome (the set of transcripts generated by the
genome of one organism) in eukaryotes and it is also involved in
the onset of several diseases~\cite{Caceres}.

A great extent of work has been performed to solve two basic
problems on AS:  characterizing the exon-intron structure of a gene
and finding the set of different transcript isoforms that are
produced from the same gene. Some computational approaches to these
crucial problems have been  proposed; indeed good implementations are
available~\cite{HeberASTP02,JeremyLeipzig08032004,Lee2,ECgene,ensamble,ASPicWeb}.
In this paper we focus on providing an algorithm -- efficient from
both a theoretical and an empirical point of view -- to predict
the exon-intron structure: a problem that is still unsolved by
methods based on  transcript data.

Indeed, it must be pointed out that few efforts have been done in
the direction of providing a formal framework to design efficient
algorithms for the general  AS prediction problem. As a matter of
fact, existing tools are not able to process  efficiently genes
that are huge or with a very large set of associated clusters of
ESTs~\cite{ASPicWeb}. A basic reason of this
fact is that combinatorial methods for the problem must combine
two different steps: (1) to produce putative spliced alignments of
ESTs against the gene region and (2) to use redundancy and the
whole covering of the gene region provided by a cluster in the
prediction process by selecting among  putative spliced alignments
of multiple ESTs~\cite{brief06} the ones that
confirm the same gene structure.
The  two steps are much harder when combined.
In fact  the literature provides efficient solutions of the first
step when
solved independently from the second one.  Finding an alignment of
an EST sequence could be a hard task when more than one alignment
can exist for the same input data and a primary goal is the choice
of the  alignment that  integrates biological meaningfulness
w.r.t.~the whole gene structure.  On the other side the second
step is NP-hard~\cite{WABI09}
thus requiring efficient heuristics.

In this paper we show how to  efficiently solve the
integration of the above two steps. First we design a new  fast
algorithm for producing splicing alignments of EST sequences by
exploiting a new combinatorial formulation of the problem.
Afterwards, given the  spliced alignments  for each EST sequence of a
gene, for each
EST we extract  a composition that is consistent with a putative exon-intron
structure of the given gene, by applying the \emph{Minimum Factorization
Agreement (MFA)} approach to the data produced by our spliced
alignment algorithm~\cite{WABI09}. Indeed, the MFA approach
provides an effective method to extract the compositions in such a way that
the whole
gene structure is derived (as an EST sequence provides the
information on a partial region of the whole gene).

\section{System and Methods}

Our new combinatorial method for exon-intron structure prediction
 can be summarized as a four-stage pipeline where we:
% \begin{enumerate}
% \item
(1)
compute and implicitly represent of all the spliced
  alignments of a transcript sequence (EST or mRNA) against a genomic reference sequence;
%\item
(2)
filter  all  biologically meaningful spliced alignments;
%\item
(3)
reconcile the spliced
  alignments of a set of correlated transcript sequences into a consensus gene
  structure;
%\item
(4)
extract, classify  and refine the resulting
  introns.
%\end{enumerate}
%
%In the following we describe all steps
%
In the following we describe the main idea on which the first two
steps of our pipeline are built.

A basic ingredient of most computational approaches for gene
structure prediction is aligning  several ESTs against the
reference genomic sequence~\cite{kan,vingron2,Xu,Galante,Xie,ES}
taking into account the effects of the excision of the intronic
regions. Thus, when considering ESTs, a particular kind of
alignment problem arises: the spliced sequence alignment. The
spliced sequence alignment problem requires to compute, given a
sequence $S$ and a reference sequence $T$,
two sets $F_S=\{f_1, \ldots, f_k\}$ and $F_T=\{f'_1, \ldots,
f'_k\}$ of strings such that $S=f_1 \cdots f_k$, $T=p f'_1 i_1
\cdots i_{k-1} f'_k s$, and for each $i$, the edit distance
between $f_i$ and $f'_i$ is small. The sequence of pairs $(f_i,
f'_i)$ is called \emph{composition} of $S$ on $T$, each factor
$f_i$ is called spliced \emph{sequence factor} (or EST factor),
and each $f'_i$ is called \emph{genomic factor} (or exon). Clearly, in the
biological context, the sequence $S$ is an EST or a mRNA
sequence, while the reference sequence $T$ is the genomic sequence
of the locus of the gene where the EST comes from. Also allowing a small
edit distance between two factors is justified by the fact that  EST data
contain mismatches (deletions and insertions) against the genome
because of sequencing errors and polymorphisms.

Notice that allowing an approximate alignment between factors
makes the spliced alignment problem computationally harder,
especially when EST data and the genomic sequence  are large.
Moreover, multiple compositions can exist for the same input data
and thus integrating biological meaningfulness is a primary goal.

One of the main ideas of our approach is that the small edit
distance between a sequence factor and a genomic factor implies
that there exist some perfectly conserved subfactors (called
pairings) between those factors. The problem of combining together
such pairings is simplified by the introduction of a new data
structure called {\em Embedding Graph} which is a graph
representing all spliced alignments. Building and querying the
Embedding Graph  is not trivial and is the main subject of our
algorithmic investigation discussed in the next section.

The whole four-step pipeline  method has been implemented  and
experimented. We designed our experimental analysis in two parts,
according to the two goals that we wanted to achieve, namely to
investigate the scalability of our implementation and to assess
the quality of the results obtained. For each gene the associated
human Unigene cluster has been processed by our spliced alignment
algorithm, together with the whole ENCODE region. Indeed, we have
assessed the quality of the results by comparing the outputs
computed by our implementation with the gene structure data
provided by ENCODE which aimed at providing a reliable annotation
of 1\% human genome~\cite{ENCODE}.

For the assessment of the prediction accuracy, we use two quality
measures, \emph{sensitivity} and \emph{specificity}, commonly adopted
for the evaluation of computational gene-structure prediction tools~\cite{Guigo2006}.
We call \emph{benchmark set} $B$ and \emph{test set} $I$
respectively the data retrieved from ENCODE and the results computed by
our approach.
Sensitivity is defined as the ratio $|B\cap I|/|B|$), while specificity\footnote{%
We  adopted the definition of specificity used when evaluating gene-structure
prediction tools, even if  not standard~\cite{DGAltman06111994}}
is defined as the ratio $|B\cap I|/|I|$.
Notice that
both ratios have values between $0$ and $1$, and that higher
values correspond to higher similarity between $B$ and $I$. We
will analyze the results produced by our pipeline according to two
coordinates: the set of predicted introns and the set of predicted
splice sites. When considering the set of predicted introns, an
intron $e\in B$ belongs to $B\cap I$ if there exists a
prediction $i\in I$, perfectly matching $e$ both on the donor
and the acceptor splice sites on the genomic sequence.

The first experiment has been run on a set
of $112$ fairly typical genes from
$13$ ENCODE regions with lengths ranging from 0.5Mb to 1.7Mb
encompassing $98,064$ transcripts (total length $63$Mb), on which we have
studied both the running times and the quality of the
results.
The second experiment, devoted to studying the scalability of our approach,
has been performed on some hand-picked genes exhibiting a
large genomic sequence or a large set of transcripts.
More precisely,  we have chosen
26 genes of which 11 are at least 1Mb long (on
average about 848Kb), and 5 have more than
15,000 transcripts (on average more than 5000 transcripts).
The complete list of regions and genes studied in both experiments, as well as
the data supporting our analysis, is in the
supplementary material.

\section{Methods}

\subsection{Implicit Computation of Spliced Alignments}

The first stage of our gene structure prediction method computes the set
of all possible spliced alignments of an EST (or mRNA) sequence
against the genomic sequence.
In particular, the first stage computes an implicit compact
representation of the spliced alignments, which is then used by the
second stage to extract all  biologically plausible alignments.

In our novel alignment method, we exploit a fundamental property of the
notion of composition: the edit distance between each pair of
corresponding factors is small.
Therefore, there must exist some  sufficiently long common substrings of the EST
factors and the genomic factors.
For example, a typical low-quality sequencing technology produces a
50bp-long exon with a 6\% error rate.
The worst scenario still allows for pairs of 11bp-long substrings that
perfectly match.
Clearly, if the sequence of perfectly matching pairs of substrings is
known, it is quite easy to reconstruct a possible alignment of the factors.
We call the sequence of the occurrences of perfectly matching pairs an
\emph{embedding} of the EST sequence in the genomic sequence.
Notice that there might be more than one embedding of the EST sequence $P$ in the genomic
sequence $T$.
We can compute efficiently all embeddings maintaining a graph whose vertices
are  all possible \emph{maximal pairings} of $P$ and $T$.
A maximal pairing of $P$ and $T$ generalizes the notion of maximal pair
of a sequence~\cite{Gusfield} and it provides the occurrence on $P$ and
$T$ of a maximal common substring of $P$ and $T$.
The vertex set $V$ can be computed in time linear in the sizes of $P$,
$T$ and $V$.
Edges of the graph are computed in time at most $O(|V|^2)$.

According to the traditional notation, given a string $S=s_1 s_2 \cdots
s_q$, we denote with $|S|$ its length and with $S[i,j]$ the substring $s_i s_{i+1} \cdots s_j$.
A fundamental notion is that of
\emph{pairing} of two strings.
Given a pattern $P$ and a text $T$, a \emph{pairing}
of $P$ and $T$ is a triplet $(p,t,l)$ such that $P[p,p+l-1]=T[t,t+l-1]$.
In other words, a pairing $(p,t,l)$  represents a common substring $x$
(or \emph{factor}) of $P$ and $T$ of length $l$ starting in positions
$p$ and $t$ on $P$ and $T$ respectively.
We call $p$ and $t$ the \emph{starting position}
on $P$ and $T$ respectively, $p+l$ and $t+l$ the \emph{ending position}
on $P$ and $T$ respectively, while $l$ is the \emph{length} of the
pairing.
When no ambiguities arise, the two strings $P$ and $T$ will be omitted when
dealing with a pairing.

Let $f$ be the common factor represented by a pairing $v=(p, t, l)$.
Notice that all triplets $(p+\delta_1,t+\delta_1,l-\delta_2)$, with $0 <
\delta_1 < l$ and $\delta_1 \leq \delta_2 \leq l - \delta_1$ are also pairings.
It is natural to define an order $\preceq$ among pairings.
Let $v_1=(p_1,t_1,l_1)$ and $v_2=(p_2,t_2,l_2)$ be two pairings, then
$v_1 \preceq v_2$ iff $p_2 \leq p_1 < p_1+l_1 \leq p_2+l_2$
and $p_1-p_2=t_1-t_2$.
When $v_1 \preceq v_2$ we say that $v_1$ is a \emph{sub-pairing} of
$v_2$, or $v_2$ \emph{contains} $v_1$.
Moreover,  $v_1$ is a \emph{prefix-pairing}
(\emph{suffix-pairing}, resp.) of $v_2$ iff $v_1\preceq v_2$
and the starting positions (the ending positions, resp.) of $v_1$ and
$v_2$ on $P$ and $T$ are equal.

Based on the order relation $\preceq$ we can define the concept of
maximality of pairings.
A pairing $v$ is \emph{maximal} if and only if it is neither a
suffix-pairing nor a prefix-pairing of a distinct pairing $v'$.
In other words, $v=(p,t,l)$ is maximal if and only if $P[p-1] \neq
T[t-1]$ and $P[p+l] \neq T[t+l]$.

We recall that an embedding is a sequence of pairings. Therefore we can extend
$\preceq$ to an order between embeddings and, based on this, we derive the
notion of maximal embeddings.
Given two embeddings $\emb=\langle v_1, \cdots, v_n \rangle$ and
$\emb'=\langle v_1', \cdots, v_m' \rangle$,  then $\emb$ is contained in
$\emb'$ (in short $\emb \preceq \emb'$) if and only if for each $v_i$
in $\emb$ there exists a pairing $v_j'$ in $\emb'$ such that $v_i
\preceq v_j'$.
Given the set $\embset$ of the embeddings of $P$ in $T$, we
say that $\emb \in \embset$ is \emph{maximal} iff there does not
exist $\emb' \in \embset$, $\emb\neq\emb'$,  such that $\emb \preceq \emb'$.

Not all embeddings induce a biologically meaningful
composition.
For example, an embedding constituted by several short pairings
``scattered'' along the genome cannot be  considered a valid spliced alignment.
A \emph{representative embedding} is a maximal embedding $\emb
= \langle v_1, \dots, v_m \rangle$ such that $l_i \geq \l_E$ and
$p_{i+1} - p_i - l_i \leq \l_D $,
and either ({\it i}) $ |t_{i+1} - t_i - p_{i+1} + p_i | \leq \l_D$
or ({\it ii}) $t_{i+1} - t_i - p_{i+1} + p_i  \geq \l_I$ is true -- only representative
embeddings   might induce a biologically plausible composition.
Intuitively, the parameter $\l_E$ is the minimum length of a pairing (in
order to avoid the previous example),  $\l_D$ regulates the
maximum number of consecutive mismatches, and $\l_I$
represents the minimum length of a valid intron.
For each composition such
that every $\l_D$-long substring of the EST sequence is aligned with the
corresponding genomic factors with up to $\frac{\l_D}{\l_E}$ errors,
there exists (at least) one representative embedding inducing such a
composition (if we choose  carefully the values of the parameters).
This fact roughly means that spliced alignments with error rate at
most $\frac{1}{\l_E}$ can be recovered from some
representative embeddings.

Therefore
% Since representative embeddings provide the information for
% reconstructing compositions of a pattern $P$ and a text $T$,
we propose
the problem of finding all the representative embeddings of
$P$ in $T$, formalized as the \textsc{Representative Embedding} Problem
    (RE), where we are given a pattern $P$, a text $T$, and three parameters $\l_E$,
  $\l_D$ and $\l_I$. The goal is to compute the set $\mathcal{E}_{r}$ of the
representative   embeddings of $P$ in $T$.
We are now able to introduce the embedding graph, which is our main device for
tackling the RE problem.
\begin{Definition}[Embedding Graph]\label{def:eg}
Given a pattern $P$ and a text $T$, the \emph{embedding graph} of $P$ in
$T$ is a directed graph $G=(V,E)$ such that the vertex-set $V$ is
composed by the set of maximal pairings of $P$ and $T$ longer than
$\l_E$, and two pairings $v_1=(p_1, t_1, l_1)$ and $v_2=(p_2, t_2, l_2)$
are connected by an edge $(v_1, v_2)$ if and only if:
% \begin{itemize}
% \item[-]
(i)
$p_2 - p_1 - l_1 \leq \l_D$, and
% \item[-]
(ii)
$|t_2 - t_1 - p_2 + p_1| \leq \l_D$ or $t_2 - t_1 - p_2 + p_1 \geq  \l_I$.
% \end{itemize}
\end{Definition}

Basically the above conditions ensure that if two maximal pairings
$v_1$ and $v_2$ are connected by an edge in the embedding graph, then
there exists a representative embedding $\emb$ in which a sub-pairing
of $v_1$ and a sub-pairing of $v_2$ are consecutive.
It is possible to prove that also the converse proposition is true:
that is, if two pairings $v'_1$ and $v'_2$ are consecutive in a
representative embedding $\emb$, then the maximal pairings $v_1$ and
$v_2$ which contain $v'_1$ and $v'_2$, respectively, are connected by
an edge in the embedding graph.
The previous properties derive from the maximality of the representative
embeddings and from the uniqueness of the maximal pairing containing a
pairing which belongs to a representative embedding.
\begin{comment}
\notaestesa{Yuri}{Inserire qualche commento per giustificare l'assenza
  della dimostrazione?}
\end{comment}

We designed an algorithm which builds the embedding graph of a pattern
$P$ and a text $T$ in time $O(|T|+|P|+|V|^2)$.
The algorithm is composed by two steps:
In the first step, the vertex-set $V$ is computed by visiting the
suffix-tree of the text $T$.
This step requires $O(|T|)$ time for the suffix-tree construction and
$O(|P|+|V|)$ time for the computation of maximal pairings.
In the second step, edges are then computed by checking the conditions
of Definition~\ref{def:eg} on each pair of maximal pairings, leading to
a $O(|V|^2)$ procedure.
Since the number of maximal pairings is usually very small compared to
the length of $P$ and $T$, the embedding graph construction procedure is
efficient even on large patterns $P$ and texts $T$.

\subsection{Extraction of Relevant Spliced Alignments}

\begin{figure}\centering
  \includegraphics[height=0.99\linewidth,angle=-90]{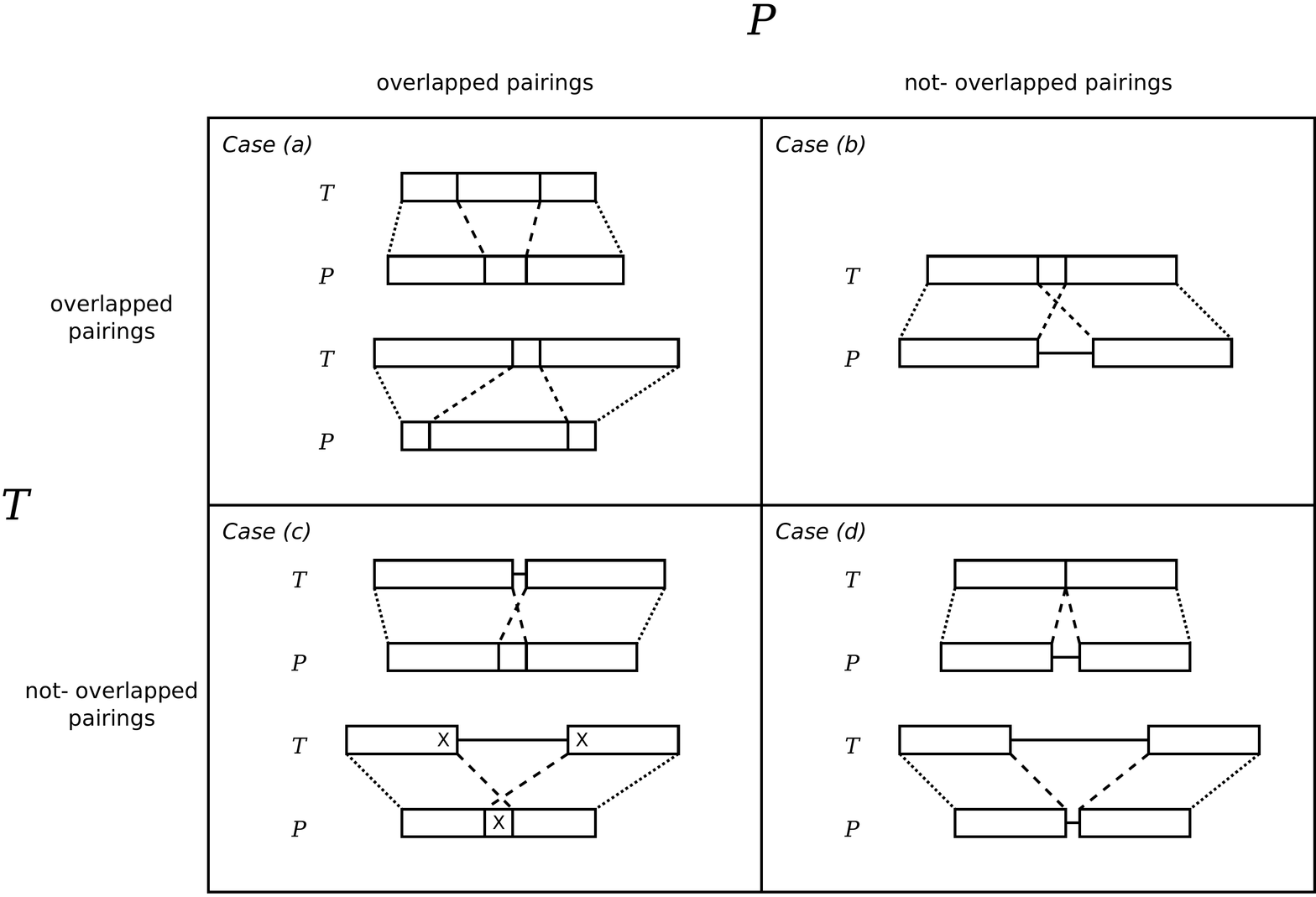}
\caption{%
Possible configurations of relative positions of two maximal
pairings $v, v'$  connected by an embedding graph edge. Each box
represents a common maximal factor on $T$ (top) and $P$ (bottom)
of a maximal pairing. Then $v, v'$  are represented by two boxes
connected by dotted lines. Four possible cases are presented: {\it
(a)} $v, v'$ overlap on both $T$ and $P$, {\it (b)} $v,
v'$ overlap on $T$ but not on $P$, {\it (c)} $v, v'$ overlap on
$P$ but not on $T$, and {\it (d)} $v, v'$ do not overlap neither
on $T$ nor on $P$.} \label{fig:edge-cases}
\end{figure}

The next stage of our pipeline is devoted to analyzing and mining the
embedding graph to compute the representative embeddings that also
induce \emph{distinct} biologically meaningful compositions.
Given two pairings that are connected by an edge in the embedding graph, the
corresponding factors might be overlapping in the text or in the pattern,
leading to four different configurations that are  depicted in
Fig.~\ref{fig:edge-cases}.

Algorithm ComputeCompositions is a two-step procedure.
Initially it extracts a subset of representative embeddings by performing a
visit of the embedding graph. Then the algorithm computes the
compositions by merging consecutive pairings that are separated by short
gaps.
Basic biological criteria (such as the recognition of canonical splice
sites) are locally used, if possible, to resolve ambiguities in the
exon-intron boundary determination.
These boundaries %, however, are not completely fixed at this stage but
can be changed during the last stage of the pipeline according to
some refinement criteria which globally analyze the set of introns induced by
the spliced alignments of all the transcripts.

\paragraph{Embedding graph visit.}
The first step of ComputeCompositions is a
recursive visit of the embedding graph starting from a subset of
vertices that we call \emph{extended sources}.
The visit of a vertex $v_k$ from the extended source $s$
reconstructs the set $\embset$ of biologically meaningful representative
embeddings that are induced by the path $\mathcal{P} = \langle s, v_1,
\dots, v_k \rangle$ traversed during the visit.
Due to space constraints we are unable to provide the
formal characterization of extended sources, as well as  some technical
details of the visit and a proof of the correctness of the step --
each  representative embedding has been examined,  only the biologically
meaningful ones have been added to set $\embset$, and  the visits compute
pairwise-distinct representative embeddings.

During the visit of vertex $v_k$, we examine each outgoing edge $(v_k, v_{k+1})$
and we ``extend'' each embedding $\emb = \langle u_1, \dots, u_k \rangle$
of $\embset$. How the extension is performed depends on the overlapping of the
involved factors, \ie it depends on the four possible configurations of
relative
positions between the last pairing $u_k$ of $\emb$ and the new vertex
$v_{k+1}$ that are depicted in Fig.~\ref{fig:edge-cases}. In the exposition of
the four cases, let  $u_k=(p_k, t_k, l_k)$ and $v_{k+1}=(p_{k+1}, t_{k+1}, l_{k+1})$.
\paragraph{Case (a).}
Factors $u_k$ and  $v_{k+1}$ overlap on both $T$ and $P$.
Two different sub-cases must be analyzed:
If $|(t_{k+1} - t_k) - ( p_{k+1}  - p_{k} )| \leq \l_D$, then the two pairings
may belong to the same factor of the induced composition.
Thus, the algorithm replaces pairing $u_k$ in $\emb$ with the shortest
maximal prefix-pairing $u'_k$ of $u_k$ and the longest maximal
suffix-pairing $u_{k+1}$ of $v_{k+1}$, such that both $u'_k$ and $u_{k+1}$ do
not overlap and have lengths at least as large as $\l_E$.
The second case is when  $(t_{k+1} - t_k) - (p_{k+1} - p_k ) \geq \l_I$. In
such case  the two pairings might be separated by an intron.
Thus, some basic biological criteria are used to compute two maximal
sub-pairings  $u'_k$  of $u_k$ and  $u_{k+1}$ of $v_{k+1}$ that could
represent a suffix and a
prefix (respectively) of an exon.
We replace $u_k$ with $u'_k$ and we extend the embedding with  $u_{k+1}$.
Whenever more than one such a pairing exists,
the embedding  $\emb$ is extended
into a set of embeddings, one for each pairing found.
\paragraph{Case (b).}
Factors $u_k$ and  $v_{k+1}$ overlap on $T$ but not on $P$.
This case is equivalent to the first sub-case of Case (a).
% , and the algorithm
% proceeds in the same way.
\paragraph{Case (c).}
Factors $u_k$ and  $v_{k+1}$ overlap on $P$ but not on $T$.
This case is similar to the entire Case (a).
Notice that when the second subcase is relevant, that is $(t_{k+1} - t_k) -
(p_{k+1} - p_k) \geq \l_I$, then the splice site placement is ambiguous because a
suffix of the donor exon is equal to a prefix of the acceptor exon.
Also in this case, basic biological criteria are used to reduce the
impact of the ambiguity.
\paragraph{Case (d).}
Factors $u_k$ and  $v_{k+1}$
do not overlap neither on $P$ nor on $T$.
Let $G_T$ and $G_P$ be the two substrings  which separate $u_k$ and
$v_{k+1}$ on $T$ and $P$, respectively. Since $G_P$ and $G_T$ do not form a
pairing, they must contain a certain number of mismatches; we must evaluate if
they support the hypothesis that (i) $u_k$ and $v_{k+1}$ are part of the same
factor or (ii) there is an intron between $u_k$ and $v_{k+1}$.
Similarly to Case (a), two different sub-cases may arise:
If $|(t_{k+1} - t_k) - (p_{k+1} - p_k)| \leq \l_D$, then $u_k$ and $v_{k+1}$ might
belong to the same factor of the induced composition. More precisely,  $u_k$
and $v_{k+1}$ belong to the same factor if
the edit distance between $G_T$ and $G_P$ is below a certain threshold -- in
which case
$v_{k+1}$ is added to embedding $\emb$, otherwise the edge is discarded from the visit.
Instead, if $t_{k+1} - t_k - p_{k+1} + p_k \geq \l_I$, the two pairings
are separated by an intron, and we must determine the splice sites of such an intron.
In this case, the algorithm computes a prefix $G'_T$ and a suffix
$G''_T$ of $G_T$, that minimize the edit distance between $G_P$ and the
concatenation of $G'_T$ and $G''_T$.
Also in this case, if the resulting edit distance is larger than an
acceptable threshold, the edge $(v_k, v_{k+1})$ is discarded, otherwise
$v_{k+1}$ is added to $\emb$. Notice that computing the edit distance is not
too expensive, since all strings involved are no longer
than $2\l_D$.

The definition of embedding graph (Def.~\ref{def:eg}) allows the
presence of directed cycles, which potentially might be troublesome.
However, cycles are of no biological interest, as no actual embeddings can be
associated to any cycle, therefore we simply ignore any edge ending in an
already visited vertex.
In fact, we claim that the embeddings, computed from a path $\mathcal{P}$
containing a cycle $\mathcal{C}$, would induce (in the following step of this
stage) compositions with essentially the same set of factors of the
compositions induced by the embeddings computed from the visit of the
simple path $\mathcal{P} \setminus \mathcal{C}$.
We support our claim with a simple example on paths containing a
directed cycle of length 2, that is the presence of a pair $(v_1, v_2)$
and $(v_2, v_1)$ of consecutive edges.
%The example can be easily generalized to cycles of arbitrary length.
Let $\mathcal{P} = \langle v_1, v_2, v_1, v_3 \rangle$ be such a path
and, for simplicity, suppose that the relative position of $v_1$ and
$v_3$ is the one depicted in the second sub-case of Case (d).
Observe that the relative position of $v_1$ and $v_2$ must be the one of
Case (a), whatever direction is considered, since a gap between the two
pairings (on $P$ or on $T$) would imply the absence of one of the
parallel edges.
Moreover, only the first sub-case of Case (a) may arise.
After the visit of path $\mathcal{P}$, the set $\embset$ of embeddings
will contain an embedding $\emb = \langle v_1', v_2', v_1'', v_3
\rangle$ where $v_1'$ is a prefix-pairing of $v_1$, $v_2$ is contained
in $v_2$, and $v_1''$ is a suffix-pairing of $v_1$.
As explained in the first sub-case of Case (a), the following step of
composition reconstruction will ``merge'' pairings $v_1'$, $v_2'$, and
$v_1''$ in a single factor (exon) since they are not separated by a gap
large enough to represent a plausible intron.
Such a factor will be approximately equal to the factor represented by
pairing $v_1$ because $v_1'$ and $v_1''$ are, respectively, a
prefix-pairing and a suffix-pairing of $v_1$.
Instead, the visit of simple path $\mathcal{P}'= \langle v_1, v_3
\rangle$ computes the embedding $\emb' = \langle v_1, v_3 \rangle$.
Clearly, also in this case, one of the factors computed in the following
step of composition reconstruction will be (approximately) the factor
represented by pairing $v_1$, concluding the example.

The visit performed in the first step of algorithm
ComputeCompositions guarantees that each possible
representative embedding is analyzed.
However, the biological criteria that we employ allow to consider only
pairings belonging to biologically meaningful embeddings.
Since the visit computes pairwise-distinct representative embeddings
and every case presented above requires $O(1)$ time, the
overall computational complexity of the visit is clearly bounded by
$O(\sum_{\emb \in \embset} |\emb|)$, that is the total size of the
representative embeddings that have been computed during the
visit.

\paragraph{Composition reconstruction.}
The set $\embset$ of representative embeddings computed by the visit of
the embedding graph directly leads to a set $C$ of compositions.
In fact, the visit guarantees that two consecutive pairings of a
representative embedding are either separated by a ``small'' gap
due to errors or by a ``large'' gap representing an intron
of the spliced alignments.
Hence, the algorithm simply merges into a factor a %(maximal)
sequence of
consecutive pairings
$v_k=(p_k, t_k, l_k)$ and $v_{k+1}=(p_{k+1}, t_{k+1}, l_{k+1})$ separated by
``small'' gaps, that is $|t_{k+1} - t_k - p_{k+1} + p_k| \leq \l_D$.
\begin{comment}
Then, the composition is reconstructed by computing an EST factor and a
genomic factor from each small-gap pairing sequence.

In particular, the EST factor (the genomic factor, resp.) is determined
as the smallest a substring of $P$ (of $T$, resp.) which includes all
the pairings of a short-gap pairing sequence.
Preliminary splice site recognition between each pair of consecutive
factors of the composition is performed by applying the same criteria
explained in Case~(d) of the previous step.
\end{comment}
Finally, the composition is retained if the edit distance between each
EST factor and the corresponding genomic factor is not greater than a
fixed acceptable threshold.

\subsection{Building a Gene Structure}

In the third stage of our pipeline, we compute a maximum-parsimony
consensus gene-structure starting from the compositions of a
cluster of transcript sequences against a common genomic sequence.
Let $T$ be the genomic sequence of a given gene locus and let $S=\{P_1,
\dots, P_k\}$ be the cluster of transcript sequences that map to the
gene locus.
The first two stages of our pipeline have  considered separately each
transcript sequence $P_i$ and, for each of them, a set $C(P_i)$ of
biologically meaningful compositions has been computed.
The main task is to extract a composition for each transcript that explains the
putative gene structure.
The redundancy of information, due to compositions of different transcripts in
the cluster, is an important ingredient to discover a single
composition of each transcript that agrees with the gene structure.
To achieve this goal we apply the \emph{Minimum  Factorization
Agreement} (MFA) problem~\cite{WABI09}.

Let us recall the definition of the MFA problem.
Let $S$ be a set of sequences over a finite alphabet $\Sigma$ of symbols,
and let $F=\langle f_1, f_2,\ldots , f_{|F|}\rangle$ be a finite ordered
set of sequences over alphabet $\Sigma$, called \emph{factors}.
Given a sequence $s \in S$, a \emph{factor-composition} (\emph{f-composition}
in short) of $s$ consists of the sequence $\langle
f_{i_1}, f_{i_2}, \cdots,f_{i_n}\rangle$ such that $s = f_{i_1},
f_{i_2}, \cdots,f_{i_n}$ and $i_j<i_{j+1}$ for $1\le j\le n-1$.
While the notion of f-composition depends on the set of factors, such set
of factors is usually clear from the context and is therefore omitted
in the definition.
Please notice that a sequence $s$ can admit different f-compositions:
thus let $F(s)$ be the set of compositions of $s$.
Moreover, by extension, we will denote by $F(S)= \cup_{s \in S} F(s)$
the set of f-compositions of a set $S$ of sequences.

Given a f-composition $f=\langle f_{i_1}, f_{i_2}, \cdots, f_{i_n}
\rangle$, the set $\{ f_{i_1}, f_{i_2}, \cdots, f_{i_n}\}$ is called the
\emph{factor set} of $f$ and is denoted as $F(f)$.
Given a subset $F' \subseteq F$ of factors and the set $F(S)$, then $F'$
is a \emph{factorization agreement set for $F(S)$} if and only if for
each sequence $s \in S$, there exists a f-composition $f$ in $F(s)$ such
that its factor set is a subset  of $F'$, i.e. $F(f) \subseteq  F'$.

The \emph{Minimum Factorization Agreement} problem, given a
set $F$ of factors and a set
$S$ of sequences, asks for a minimum cardinality subset
$F' \subseteq F$ such that $F'$ is a factorization agreement
set for $F(S)$.
Informally, the solution to the MFA problem is a smallest  set of factors that
is able to explain a f-composition for each input sequence.
In our setting
%Now assume that
$S$ is  the cluster of transcript sequences and $F$ is
the set of all exons (factors) used to produce the compositions of sequences
in $S$, i.e. $F(S)$ consists of all the compositions of $S$.
When solving the MFA problem on these input data, the solution $F'$ provides
a minimum set of factors that can explain all transcript sequences and a single
composition of each transcript can be obtained from set $F'$.

However, in order to apply the MFA problem to our data we need to
define a binary relation between factors (exons).
Indeed,  the first and the last factors of different transcript sequences could
be fragments of the same exon.
Moreover, since internal factors of transcript compositions are computed
without applying refined biological criteria for the location of splice
junctions, even internal factors of different transcript compositions could be
associated to the same exon whenever they differs by few bases.
For this reason we define a binary relation between factors of
compositions.
More precisely, we say that two factors $f$ and $f'$ are $\sim$-related
if the two factors share a common overlapping regions of length at least
$N$, for $N$ a fixed bound.
Factors that are $\sim$-related are grouped into a \emph{region}:
each composition $c$ of a transcript sequence $s$ is then replaced by a
composition into regions corresponding to each factor (recall that
regions are disjoint).
Similarly the set $F$ consists of regions and thus the MFA problem
applied to the new set $F(S)$ produces a minimum set $F'$ of regions
that could explain all input compositions.

By applying the algorithm proposed by~\cite{WABI09} we are able to
filter efficiently a set of spliced alignments agreeing to the same gene
structure that are further refined by the next step of intron reduction.

\subsection{Intron Reduction}
Although the intron boundaries of EST spliced compositions are computed by
finding the best transcript-genome alignment over the splice site
regions and the most frequent intron pattern (\ie the first and the
last two nucleotides of an intron) according
to~\cite{Burset1}, the set of predicted introns may still contain
false positives very close to true predictions.
\begin{figure}
  \centering
  \includegraphics[width=0.98\linewidth]{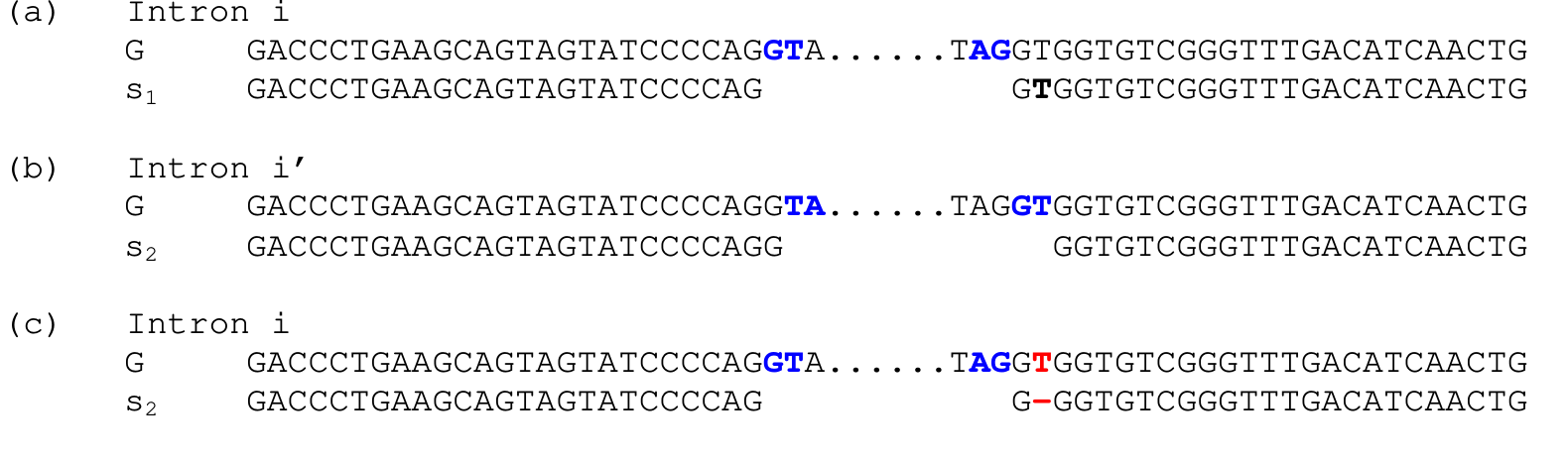}
  \caption{An example of intron reduction}\label{fig:IRMotivation}
\end{figure}
This can be explained by the example in
Figure~\ref{fig:IRMotivation}. Let us suppose that there exists an
EST $s_1$ (see Figure~\ref{fig:IRMotivation}(a)) producing a
canonical intron $i$ -- \ie satisfying the so-called $GT-AG$
rule~\cite{Burset1} --  and
that $s_1$ perfectly matches to the genomic sequence over the
intron splice sites. Moreover, let us suppose that there exists
another EST $s_2$ (see~Figure \ref{fig:IRMotivation}(b)) with a
deletion of a base 'T' with respect to $s_1$ (highlighted in boldface
in Figure~\ref{fig:IRMotivation}(a)). Because of our spliced
alignment algorithm, $s_2$ produces a non-canonical intron $i'$
(indeed its pattern is $TA-GT$), which differs from $i$ by only
a few bases. We can reasonably assume that $i'$ is a false positive
and not a reliable prediction. Hence, the spliced composition of
$s_2$ can  be corrected into the spliced alignment
supporting the intron $i$ (see Figure~\ref{fig:IRMotivation}(c)),
by introducing just one insertion error into the genomic sequence.
Thus, a procedure for comparing the intron set computed by the
EST spliced compositions in order to correct and reduce the set of
false positives (i.e. over predicted introns) is necessary.

In the following, let the pair $(i,s)$   denotes a genomic  intron
(eventually specified by a pair of genomic coordinates) and a
spliced composition  of an EST $s$ supporting  the intron $i$,
i.e. having two consecutive factors $f_i$, $f_{i+1}$ inducing
intron $i$ when aligned to the genome. Then, given an error bound
$b$, we say that $(i,s)$ is $b$-reducible to $(i',s)$ iff there
exists a boundary  shift of factors $f_i$ and $f_{i+1}$ of a new
spliced composition of $s$ inducing  intron $i'$ with at most
additional $b$ errors w.r.t. the previous alignment of the two
factors against the genome. Since  RefSeq transcripts are usually
full-length and error-free, and $GT-AG$, $GC-AG$ and $AT-AC$ rules
are the most frequent~\cite{Burset1} and are associated to
U12/U2 introns~\cite{Sheth2006}, we assume that only introns,
that do not follow one of the U12/U2 rules and are not supported
by a RefSeq transcript should be reduced. The input of our
intron-reduction procedure  is a set $X$   of pairs $(i,s)$
computed by the previous steps. Then,
 $R$ is the set of pairs in $X$  such that
$s$ is a RefSeq,    $C_1$, $C_2$, $C_3$ and $N$  are the set of
pairs in $X \setminus R$ following the $GT-AG$, $GC-AG$, $AT-AC$ and a
non-U12/U2 rule respectively.   Our procedure basically tries to
reduce elements in $N$ to some intron in  $R$ (i.e. the most
reliable class composed by RefSeq-supported introns) and if this
is not possible tries to reduce to some element in the first set
of the sequence $C_1$, $C_2$ , $C_3$ that allows the reduction.

\section{Implementation}

We implemented the pipeline as a set of C programs
in the software package PIntron.
The input is a genomic sequence $G$ and a set $T$ of EST and
mRNA sequences, referred in the following with the more general term of transcripts.
Our method computes a set $C$ of exon compositions of the sequences in $S$
with respect to the genomic sequence, such that there exists one exon
composition per input transcript. In fact, for each transcript, our software
retains only those compositions that can possibly provide information about
the exon-intron structure.
Our implementation is still at a preliminary stage; in fact only a few basic
biological criteria  have been introduced, such as constraints on the intron
lengths as well as on the size of removable transcript prefixes and suffixes.
The main goal of our study -- and of our current implementation -- is to
assess the efficiency of the pipeline while maintaining a reasonable quality
of the results produced.
% applied, such as introns must be of length greater than a fixed value (that is
% $?$ bp in the experimentation described in the following), the coverage of a
% composition, with respect to the total transcript length, must be greater than
% a fixed threshold, the transcript prefix/suffix that can be discarded from a
% composition must be limited (in the experimentation to a $?$\% of the
% transcript length) and ... \nota{qualcosa sulla procedura di raffinamento
% dell'introne {\`e} gi{\`a} stato detto prima? E sul filtraggio in base alla
% complessit{\`a} degli esoni?}.
Together with the exon compositions, PIntron outputs
% The output of the PIntron software is the set of exon compositions of all the
% input transcripts with respect to the genomic sequence and the list of introns
% induced by the exon compositions on the genomic sequence. In particular, for
% each gene our software produce an output file containing a row for each
% predicted intron. Each row is composed of $19$ tab-separated fields reporting
% some information such as
the positions, on the genomic sequence, of the donor and the acceptor splice sites
of the introns, as well as several additional informations included
% \emph{le coordinate rispetto al cromosoma
%   mancano...},
% the list of the gb IDs of the supporting transcripts, the donor
% and acceptor splice site scoring, the scoring and the position of the Branch
% Point Sequence BPS,
the intron type (U12, U2 or unclassified)~\cite{Sheth2006}.
% the intron pattern and the sequences of the regions over
% the two splice sites.

\section{Discussion}
\label{sec:experimentation}

The results of the experimental analysis are very positive, especially the
running times, as the analysis of
several genes has required only a few hours, even
using modest computational resources. After a first experiment,
where we have analyzed a set of 112 fairly typical genes from
13 ENCODE regions with lengths ranging from 0.5Mb to 1.7Mb
encompassing 98,064 transcripts (with total length 63Mb), we have decided to
investigate the scalability of our
implementation by analyzing some hand-picked genes exhibiting a
large genomic sequence or a large set of transcripts.
Comprehensive data regarding the  experiments are
reported in the supplementary material.

In the first experiment, for each gene
the associated human Unigene cluster (including also RefSeq mRNAs)
has been processed by our spliced
alignment algorithm, together with the whole ENCODE region.
% The total length
% of the processed genomic sequences has been of 8,358,447 bp and the total
% number of processed transcripts for the 112 genes has been of 98,064, with
% an average of 876 transcripts for each gene. The total length of the
% transcripts processed has been of 63,127,862 bp.
We tested our software on
an off-the-shelf PC
with a total running time of 1h 17sec (on average 41 seconds/gene).
%Computational time results are summarized in ??.

\begin{comment}
%Spostato alla fine
An analysis on the running times has not shown any significant
correlation between the length of the genes and the running times,
hence confirming our conjecture that the behaviour of our
algorithm is highly dependent on some properties of the embedding graph,
%\nota{P:CMEG, ricordarsi che non {\`e} stato denominato CMEG per ora}
and not on the usual size of the instance. %\nota{P:
In particular,
the structure of the embedding graph is strictly related to the quality of
the transcript and to the presence in the gene of repetitions,
highly duplicated regions or other elements that could influence
the size of the graph.
%}
\end{comment}

\begin{figure*}[t!]\centering
\includegraphics[width=0.95\linewidth]{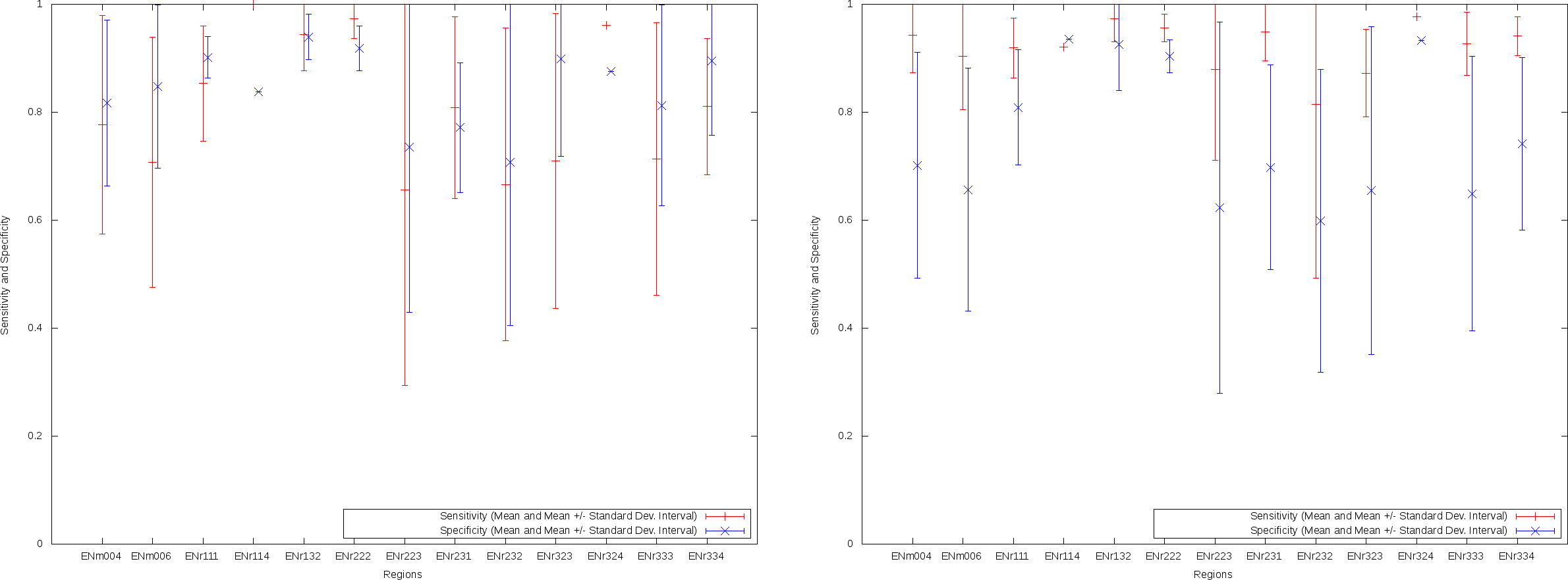}
\caption{Distribution of ENCODE introns (left) and splicing sites
  (right) confirmed by PIntron.
  For each region we have represented the mean of sensitivity and
  specificity values, together with the interval ranging from the mean
  minus the standard deviation and the mean plus the standard deviation.}
\label{fig:grafico_completo}
\end{figure*}%
\begin{figure*}[t!]\centering
\includegraphics[width=0.95\linewidth]{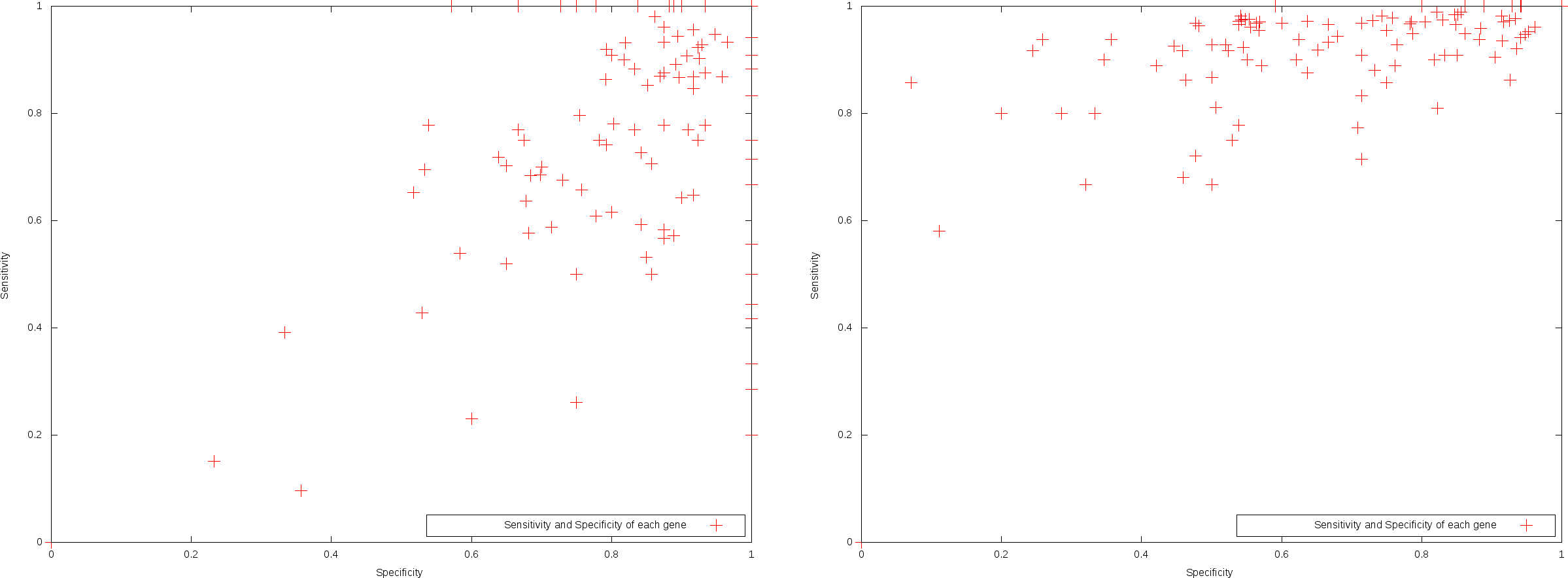}
\caption{Distribution of ENCODE introns (left) and splicing sites
  (right) confirmed by PIntron.
  Each gene is represented by a point whose coordinates are its
  specificity and its sensitivity.}
\label{fig:scatter_completo}
\end{figure*}%
On the whole set of the 112 input genes, 1787 out of 1957 benchmark
introns are also in the test set with an average support of 214.45
transcripts per intron (not including outlier  genes RPL10 and EEF1A1 that
have 6,414 and 49,936 transcripts respectively).
\begin{comment}
, which are
the largest input Unigene clusters among the experimented genes).
\end{comment}
We have further investigated the 170  ENCODE introns that have not been
predicted by our method, searching in the ENSEMBL Genome Browser the
supporting transcripts of each unpredicted intron. The investigation pointed out that
95 introns (55.88\% over the total of missing data) have no
evidence in the ENSEMBL database, hence they are probably
hand-annotated introns. Of the remaining 75 unpredicted introns,  41
(24.12\%) have no supporting transcripts in the Unigene set fed to our
implementation. 18 unpredicted introns (10.59\%) have some supporting
transcripts, but  with only low-quality alignments between each transcript and
the unpredicted intron. For most such transcripts  our method computes a
high-quality alignment supporting another ENCODE intron that is very close to the
unpredicted one.
%\nota{che non si possono allineare secondo l'introne a meno di una
%  qualit{\`a} di allineamento pessima, e che supportano nella maggior parte dei
%  casi altri ENCODE molto vicini e quindi altri TPs, oppure supportano
%  l'introne inteso come siti di splicing ma ci piazzano dentro un esone}
Finally, for 16 (9.41\%) unpredicted introns our method is unable to
align the transcripts supporting such introns (in one case we found  errors in the Unigene DB).
%(in some cases the detailed
%investigation has found  errors in the Unigene DB or has questioned the
%correctness of the unpredicted ENCODE intron).
%\nota{... e non {\`e} sempre
%  colpa nostra... 1 ha un trascritto per cui lo strand indicato da Unigene
%  {\`e} sbagliato, un altro in realt{\`a} lo troviamo a meno di poche basi,
%  semplicemente perch{\`e} l'ENCODE non {\`e} canonico e noi ne troviamo uno
%  vicino che {\`e} sempre non-canonico ma il suo pattern ha una frequenza di
%  Burset maggiore, 3 falsi negativi li troviamo in fase di fattorizzazione ma
%  poi in intron agreement li modifichiamo per raggiungere un intron pattern
%  migliore (ammettiamo forse troppi errori...), e per 11 non riusciamo proprio
%  ad allineare e quindi {\`e} molto probabile che sia colpa nostra, quindi
%  quelli per cui {\`e} colpa nostra sono in tutto 11+3=14}.
The overall sensitivity of our method is 0.9813, computed after discarding from the benchmark
set the 136 ENCODE introns that have no evidence in ENSEMBL DB or have no supporting
transcripts in the Unigene cluster given in input to PIntron. The benchmark
set contains no intron for  F8A1, H2AFB1 and IER5L.
Our method confirms ENCODE for the first two genes, while it predicts 3 introns for IER5L.
Notice that all genes (except OPN1MW) have sensitivity at least 0.8
and almost 75\%  have sensitivity 1.

Our experiment has predicted 1042 introns that are not
in the benchmark. The overall specificity of our method is 0.6317 and 10 genes
have specificity equal to
1.  Moreover, the average number of supporting transcripts for
each new intron is 2.42. 268 new predictions have one
of the two splice sites in the benchmark set $B$, while 92 new introns have both
splice sites in $B$ (but they belong to different ENCODE introns).
Sensitivity and specificity for this analysis are represented in
%Figures~\ref{fig:grafico_introns1},~\ref{fig:scatter_introns1},~\ref{fig:grafico_sites1},~\ref{fig:scatter_sites1}.
Figures~\ref{fig:grafico_completo} and~\ref{fig:scatter_completo}.

We have also studied if the introns predicted by PIntron are close to those
predicted by ENCODE, relaxing the definition of introns common to the
benchmark and test sets by allowing splice sites that differ for up to 8
positions. We omit the results as no significant improvement has resulted.

Since our pipeline seems to overpredict with respect to ENCODE, we have
started a detailed analysis to understand the causes of such overpredictions.
Almost all introns predicted by PIntron and ENCODE (1778 out of 1787) are
canonical, classified as
U12/U2 and contain a Branch Point Sequence (BPS).
Since we have not completed our analysis yet, we are currently unable to
determine if being canonical, classified or containing a BPS should be a
requirement for a putative intron to be retained in the results output by
PIntron. For example, introducing the criterion that only canonical and
classified are output, the number of new predictions decreases from 1042 to
565: in such case the sensitivity becomes 0.9764 and the specificity
becomes 0.7598.

%La frase seguente mi sembrave troppo forte, riformulata
% The only comparison among gene prediction tools in literature~\cite{BPRMJC}
% is performed over different genes, therefore that analysis and ours are not
% completely comparable. Anyway, for illustrative purposes we report here that
% the two most used software packages for
% predicting the exon-intron structure, namely ECgene~\cite{ECgene} and
% ASPic~\cite{ASPicWeb} have sensitivity/specificity that is respectively
% 0.92/0.63 and 0.88/0.77, therefore pointing our the quality of the results of
% our pipeline.
Other comparisons among gene-structure prediction tools in literature~\cite{BPRMJC}
are performed over different datasets, therefore that analysis and ours are not
completely comparable. Anyway, for illustrative purposes we report that
two commonly used software packages for
predicting the exon-intron structure, namely ECgene~\cite{ECgene} and
ASPic~\cite{ASPicWeb}, have sensitivity/specificity that is respectively
0.92/0.63 and 0.88/0.77~\cite{BPRMJC}: not as good as those obtained by our pipeline.
\begin{comment}
\notaestesa{R}{In ECgene la sensitivit{\`a} rilevata nel nostro
lavoro di JCB {\`e} stata di 0.92 (aspic 0.88) e la specififict{\`a}
(aspic 0.77) di 0.63. Inoltre prediciamo in pi{\`u} 25 nuovi introni che
non sono canonici ma sono classificati e tutti con BPS, ma forse per dire se sono buoni
occorrerebbe un'analisi un po' pi{\`u} approfondit{\`a} sul numero di EST di supporto e
sull'errore di allineamento... In sintesi dei 1042, 477 li possiamo filtrare, 540 sono
canonici e i rimanenti 25 sono non canonici ma classificati. Chiaramente in tutto ci{\`o} manca
un'analisi sul numero di EST di supporto sull'errore di allineamento e sulla presenza della BPS}
We are planning to complete the analysis of the causes for overprediction, in
order to enlarge the criteria employed by our implementation to detect the
exon-intron structure.
\end{comment}
Another parameter that is surely interesting is the
number of EST supporting a exon/intron.

%\nota{P:Sensitivity e specificity devono essere date anche in
%funzione delle osservazioni fatte su alcuni introni, cio{\`e} vanno
%tolti gli introni in Encode che non hanno evidenza per poter
%essere rilevati da PIntron.}

%We carried out also an analysis over the sets of the 1787 true positives in
%order to find some criteria and to associate to each one of the 1042
%candidate novel introns a quality score (??) and asses its
%reliability.\notaestesa{GDV}{Immagino che qui sia da completare il paragrafo}
%\nota{P:Infine sulla specificit{\`a} andrebbe data
%rispetto ad introni con certe caretteristiche, canonici o
%classificati... In ECgene la sensitivit{\`a} rilevata nel nostro
%lavoro di JCB {\`e} stata di 0.92 (aspic 0.88) e la specififict{\`a}
%(aspic 0.77) di 0.63.}

The time performance of PIntron over the experimented ENCODE genes is detailed
in the supplementary material. In this section we will show the results only for three genes
that are critical because of the size of the input transcripts  or because of
the length of the input genome (\ie the corresponding ENCODE region). We will
report only the time for computing spliced compositions, since it is the most
time-consuming part of our  pipeline.
More precisely, gene TIMP3 has taken 48 seconds for computing spliced compositions of 1,700
transcripts against the ENCODE region ENm004 that is 1,700,000 bp long. Gene RPL10 (in region
ENm006) has taken 141 sec for aligning 6,414 transcripts against a genome of 1,338,447bp, and finally
the time, for computing the spliced compositions of the 49,936 transcripts of gene
EEF1A1 against the 500,000bp of region ENr223, has been 416 sec.

\begin{comment}
\begin{figure}\centering
\includegraphics[width=0.98\linewidth]{grafico_introni1}
\caption{Distribution of ENCODE introns confirmed by PIntron. For each region
  we have represented the mean of sensitivity and specificity values,
  together with the interval ranging from the mean minus the standard
  deviation and the mean plus the standard deviation.}
\label{fig:grafico_introns1}
\end{figure}

\begin{figure}\centering
\includegraphics[width=0.98\linewidth]{scatter_introni}
\caption{Distribution of ENCODE introns confirmed by PIntron.
  Each gene is represented by a point whose coordinates are its
  specificity and its sensitivity.}
\label{fig:scatter_introns1}
\end{figure}

\begin{figure}\centering
\includegraphics[width=0.98\linewidth]{grafico_siti}
\caption{Distribution of ENCODE splicing sites confirmed by PIntron. For each region
  we have represented the mean of sensitivity and specificity values,
  together with the interval ranging from the mean minus the standard
  deviation and the mean plus the standard deviation,}
\label{fig:grafico_sites1}
\end{figure}

\begin{figure}\centering
\includegraphics[width=0.98\linewidth]{scatter_siti}
\caption{Distribution of ENCODE splicing sites confirmed by PIntron.
  Each gene is represented by a point whose coordinates are its
  specificity and its sensitivity.}
\label{fig:scatter_sites1}
\end{figure}
\end{comment}

To
% study In order to show the time performance of our method, as well as
% to
determine the scalability of our approach, we have chosen
26 genes, of which 11 are at least 1Mb long (on
average about 848Kb), and 5 have more than
15,000 transcripts (on average more than 5000 transcripts).
We tested our software on a commercial workstation equipped with 12GB of RAM
with a total running time of 20 min (average 47 seconds/gene).
%The detailed table about these genes is reported in the supplementary material.
Notice that processing TTN gene has taken  545
seconds (while other tools, such as the aforementioned ECgene, do not
provide predictions for this gene).
%\notaxestesa{GDV}{Quanto impiegano gli altri programmi sulla titina?}
The likely reason is  that the input set contains transcripts (ESTs and mRNAs)
that are more than 80Kb long, and the spliced alignment computation
time is very high for long EST sequences since their quality is lower
than the quality of mRNA sequences.

An analysis on the running times from both experiments has not shown any significant
correlation between the length of the genes and the running times,
hence confirming our conjecture that the behaviour of our
algorithm  depends on some properties of the embedding graph,
and not on the size of the instance.
In particular, the structure of the embedding graph is strictly related to the quality of
the transcripts and to the presence in the gene of repetitions,
highly duplicated regions or other elements that could influence
the size of the graph. The results have confirmed %both
our beliefs, since the average running time of the second experiment (47
seconds/gene), albeit on a faster PC,
is not too far from the running times on the much smaller genes of the
first experiment, where the average value is 41 seconds/gene.
The experimental results for all  investigated genes are available
 at \coll{http://www.algolab.eu/PIntron}.

\bibliographystyle{plain}
\bibliography{abbreviations,books,biology,complexity,splicing,stats}

\end{document}